\documentclass[12pt]{iopart}
\usepackage{graphicx}
\usepackage{amssymb}
\usepackage{lscape}
\begin{document}

\title{Role of vacuum fluctuation forces in thin metal film stability}
\date{\today}

\author{A.~Benassi$^{1,2}$, C.~Calandra$^2$}

\address{$^1$ CNR/INFM-National Research Center on nanoStructures and bioSystems at Surfaces
(S3), Via Campi 213/A, I-41100 Modena, Italy}
\address{$^2$ Dipartimento di Fisica, Universit\`a di Modena e Reggio Emilia, Via Campi 213/A, I-41100 Modena,
Italy} \ead{benassi.andrea@unimore.it}

\begin{abstract}
Thin metal films are subject to the pressure caused by the zero point oscillations of the electromagnetic field, which depends upon the film optical properties and, in case of deposition onto a substrate, upon the substrate reflectivity. It has been suggested that this force may be relevant in determining the stability of deposited pseudomorphic films with respect to buckling or island formation. We present a detailed analysis of its behaviour as a function of the optical parameters and of the film thickness and we illustrate the conditions under which it may play some role. For free standing films it turns out that the film stabilization is basically due to the surface stress, which largely overwhelms the vacuum force. For epilayers on metal substrate the vacuum force may be important, and we give stability diagrams and critical thicknesses for several cases, illustrating how the flat surface growth may depend upon the film parameters. The importance of including retardation effects into the theory for a realistic determination of the stability conditions is also discussed. 
\end{abstract}


\section{Introduction}
\label{introduction} 
Epitaxial deposition of thin metal films on various substrates is of crucial importance in several applications ranging from the design of systems for heterogeneous catalysis and electron emission to the technology of semiconductor devices and magnetic recording \cite{chopra,matthews,dreysseand,fu}. In several cases a significant mismatch between the lattice parameters of the deposit and the substrate is present. If the mismatch is not too large, the overlayer grows in a way that the atoms of the deposit are in registry with the substrate atomic structure and the growth is pseudomorphic. As the film thickness increases, mechanical strain energy builds up in the deposited film, which eventually can cause instability unfavorable to uniform flat film growth: the formation of dislocations or the continuation of growth in the form of islands become energetically favorable. In several cases epitaxially deposited film grown layer by layer can undergo a transition to a growth mode of three dimensional dislocation-free islands, which form on top of a layer of a certain critical thickness
\cite{jang,bittner,dietterle,randler,freund}. It turns out that the critical thickness is material dependent and decreases with increasing the lattice mismatch strain \cite{zhigang,kraft}.\\
The structure and morphology of the deposited films is often determined by kinetic constraints rather than by the minimization of the free energy. However it is possible to work in experimental conditions that allow the system to relax toward the equilibrium configuration. In such cases equilibrium considerations may be useful to understand the physical interactions that are responsible of the film evolution and to establish the conditions for the existence of a critical thickness and its value.\\
One can discuss the stability issue using a continuum model \cite{asaro,grienfield,stolovitz,gao2,spencer,mueller}. The value of this approach lies in the fact that one can draw qualitative and some quantitative conclusions on general aspects of the growth without the need of a microscopical description of the atomistic mechanisms present in the process. Within this model the stability of epitaxial films results primarily from the competition between the surface energy and the elastic energy, caused by lattice mismatch, since when a flat film surface is modified into a wavy shape the surface energy increases, but the elastic energy decreases. Roughening of  the surface and island formation can occur when it is energetically favorable to relax the elastic energy by increasing the surface area. 
It has been observed that since the surface energy tends to stabilize perturbations at low wavelength, while the elastic energy amplifies the perturbation at all the wavelength, any strained film, regardless of its thickness, should be unstable. This prediction does not match with the experiments which show the existence of a critical thickness beyond which the instability occurs. This suggests that some additional force has to be present that competes with elastic stress and stabilizes the film within a range of thickness of the order of few nanometers. Long range forces, in particular the vacuum fluctuation forces between the interface and the film free surface, have been considered as a possible source of stability in the system \cite{zhigang,jiang}. Due to their peculiar size dependence, these forces are expected to play a role only at nanometric distances and when they cause a repulsion between the films boundaries.\\
In this paper we discuss the conditions under which the vacuum fluctuation forces can contribute to deposited metal film stability. Our purpose is to identify, both for the film and for the substrate, the range of optical parameters, that can determine the existence of a finite critical thickness below which the flat surface growth takes place.
To this end we need a realistic description of the metal optical properties. Recently we have reported on the results of a study of vacuum electromagnetic forces on thin metal films based on the plasma model \cite{benassi}. We have shown that the force may change in sign and intensity depending upon the film size and the nature of the substrate. 
However the plasma model does not provide an accurate description of the real metal dielectric properties, since it neglects both the intraband absorption of the free electrons and the effects on the interband transitions in the relevant frequency range.\\ 
A more accurate description is give by the Drude model 
\begin{equation}
\epsilon(\omega)=1-\frac{\Omega_{p}^{2}}{\omega(\omega+i\omega_{\tau})}
\label{drude}
\end{equation}
which includes the parameters $\Omega_{p}$, the plasma frequency, and $\omega_{\tau}$ the relaxation frequency. For $\omega_{\tau}=0$, corresponding to infinite relaxation time, we recover the plasma model. While in the 
plasma model $\Omega_{p}$ is fixed by the valence electron density, the Drude model allows to extract the parameters 
from the experimental data, thus taking into account both the real electron density and the effective mass of the 
electron in the metal under consideration. Recently a careful study of the fitting procedure for Au films, focused 
on the frequency range that is relevant in the determination of the forces due to vacuum electromagnetic field 
fluctuations, has shown that the Drude model can be useful to reproduce optical experimental data for films prepared 
in various experimental conditions \cite{pirozhenko}, thus allowing for a more accurate evaluation of the force.\\
We start our analysis with a study of the force on isolated metal films. A free standing metal film is subject to 
a negative pressure i.e. an attractive force per unit area $F$ between the boundaries caused by zero point 
oscillations of the quantized electromagnetic field.
We have performed calculations of the vacuum induced electromagnetic force at $T= 0^{\circ}K$ for free standing films of different thickness using the dielectric function of equation (\ref{drude}) and analyzed its dependence upon the optical film parameters. Here we report on the results and on related theoretical aspects concerning the role of this force in the film stability. 
The study of the forces on deposited films has been carried out in two steps. First we have considered the force on a metal film deposited onto a perfectly reflecting substrate. In this case the force between the film boundaries, i.e. the substrate-film interface and the film surface, is always repulsive, independently of the film parameters, and the discussion on the stability issue, carried out with reference to standard values of the stress concentration and the surface energy, is particularly simple. We identify the range of parameters that permit the existence of a critical thickness in this ideal case. 
Turning to real metal substrates we find, in agreement with previous work based on a plasma model description \cite{benassi}, that the force can be repulsive or attractive, depending upon the the difference in the optical parameters of the substrate and the film. An attractive force contributes to the destabilization of the flat surface growth and favors the island formation. In case of a repulsive force we provide examples of stability diagrams, which give the range of parameters that allow the system to remain stable, and we plot the corresponding critical thicknesses.
Our purpose is not to derive results for a specific system, rather we illustrate the conditions under which this force may be important.
\section{Free standing metallic film}
\label{freefilm} 
\begin{figure}
\centering
\includegraphics[width=3cm,angle=0]{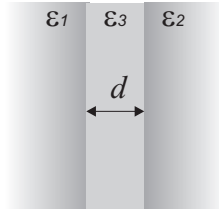}
\caption{\label{refframe3} Notation for three layers system}
\end{figure}

The expression of the vacuum fluctuations energy at $T=0^{\circ}$K in a configuration with two semi-infinite media of dielectric functions $\epsilon_{1}(\omega)$ and
$\epsilon_{2}(\omega)$ separated by a film of thickness $d$ and dielectric function $\epsilon_{3}(\omega)$ (see fig.\ref{refframe3}) is given by the Lifshitz formula \cite{lifshitz}.
\begin{equation}
E(d)=\frac{\hbar L^2}{4 \pi^2 c^2}\int_{1}^{\infty}p dp\int_{0}^{\infty}\xi^2\bigg\{ln\big[Q_{TM}(i\xi)\big]+ln\big[Q_{TE}(i\xi)\big]\bigg\}d\xi
\label{force}
\end{equation}
where:
\begin{equation}
Q_{TM}(i\xi)=1-\frac{(\epsilon_{1}K_{3}-\epsilon_{3}K_{1})(\epsilon_{2}K_{3}-\epsilon_{3}K_{2})}{(\epsilon_{1}K_{3}+\epsilon_{3}K_{1})(\epsilon_{2}K_{3}+\epsilon_{3}K_{2})}e^{-2\xi K_{3}d/c}\\
\label{qtm}
\end{equation}
\begin{equation}
Q_{TE}(i\xi)=1-\frac{(K_{3}-K_{1})(K_{3}-K_{2})}{(K_{3}+K_{1})(K_{3}+K_{2})}e^{-2\xi K_{3}d/c}
\label{qte}
\end{equation}
refer to the contribution of transverse magnetic ($TM$) and transverse electric ($TE$) modes respectively. Here
\begin{equation}
K_{i}(i\xi)=\sqrt{p^2-1+\epsilon_{i}(i\xi)}=\frac{c}{\xi}\gamma_{i}
\end{equation} 
where
\begin{equation}
\gamma_{i}(i\xi)=\sqrt{k^{2}+\frac{\xi^{2}}{c^{2}}\epsilon_{i}(i\xi)}\qquad k^2=\frac{\xi^2}{c^2}(p^2-1)
\end{equation}
$k$ being the component of the field wave-vector parallel to the planar interface. 
The force per unit area is given by:
\begin{equation}
F(d)=-\frac{1}{L^2}\frac{\partial E}{\partial d}
\end{equation}
The case of a free standing metal film is obtained with $\epsilon_{1}(\omega)=\epsilon_{2}(\omega)=1$ and $\epsilon_{3}(\omega)$ given by the Drude model.
\begin{equation}
\epsilon_{3}(\omega)=1-\frac{\Omega_{3}^{2}}{\omega(\omega+i\omega_{\tau})}
\end{equation}
where $\Omega_{3}$ is the plasma frequency of the film.
One can easily verify that the use of the Drude model leads to a vanishing force in the two limits $\Omega_{3}\rightarrow0$ (infinitely diluite metal) and 
$\Omega_{3}\rightarrow\infty$ (ideal metal). Since the force cannot be identically zero for finite plasma frequency values, we expect that it has extrema in between these two limits.\\
\begin{figure}
\centering
\includegraphics[width=8cm,angle=0]{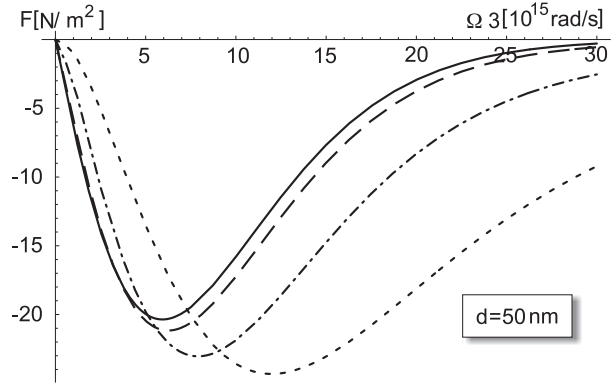}
\caption{\label{fig2}Free standing film. Force on film boundaries as a function of film plasma frequency for different relaxation frequencies. $\omega_{\tau}=0$ continuous line, $\omega_{\tau}=10^{14}rad/sec$ long dashed line, $\omega_{\tau}=10^{15}rad/sec$ dot dashed line, $\omega_{\tau}=5\cdot 10^{15}rad/sec$ short dashed line.}
\end{figure}
Fig.\ref{fig2} displays the calculated force versus plasma frequency curves for a $50 nm$ film. We have chosen $\omega_{\tau}$  comparable with those obtained
by fitting optical properties and from sample metal resistivity data \cite{ashcroft}. For comparison we give in the same figure the curve obtained using the plasma model.
It is seen that the general shape of the curve is the same: the force is attractive and shows a maximum of intensity (a minimum in the curve) and a long tail at high
frequencies. However the inclusion of a finite relaxation frequency leads to significant modifications in the force intensity both at low and high frequency and causes a shift in the position of the maximum.\\
\begin{figure}
\centering
\includegraphics[width=8cm,angle=0]{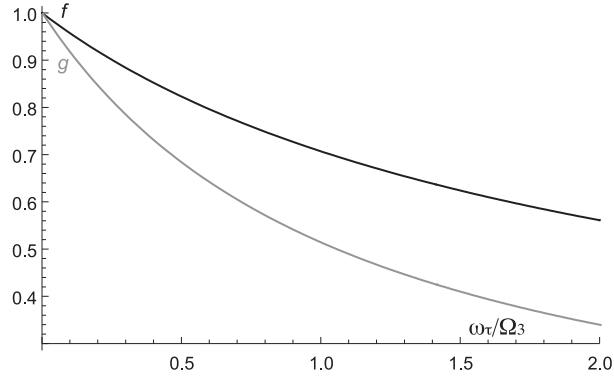}
\caption{\label{fig3}$g$ factor of equation (\ref{prodbis}), gray line; $f$ factor of equation (\ref{prod2bis}), black line.}
\end{figure}
In particular we notice that, on increasing the relaxation frequency, the force decrease at low $\Omega_{3}$ while it becomes significantly higher at high plasma frequency. The weakening of the force in the low plasma frequency range, can be understood using the small distance approximation to the Lifshitz formula, that can be applied when the film thickness $d$ is much smaller than both the plasma wave length $\lambda_{p}=2\pi c/ \Omega_{3}$ and the relaxation wavelength $\lambda_{\tau}=2\pi c/ \omega_{\tau}$. In such conditions the force is due to $TM$ modes only and is simply given by:
\begin{equation}
F(d)=-\frac{\hbar}{8 \pi^2 d^3}\int_{0}^{\infty}\frac{(\epsilon_{3}(i\xi)-1)^2}{(\epsilon_{3}(i\xi)+1)^2}d\xi
\label{dsmall}
\end{equation}
showing the typical $d^{-3}$ behaviour of the van der Waals dispersion forces. Inserting the Drude dielectric function into this equation we get:
\begin{equation}
F=F_{P1}\: g\Bigg(\frac{\omega_{\tau}}{\Omega_{3}}\Bigg)
\label{prod}
\end{equation}
where
\begin{equation}
g(x)=
\frac{\sqrt{2}}{\pi} \Bigg[\frac{2(1-x^2)-2}{x(1-x^2)}-\frac{1}{(1-x^2)^{3/2}}\bigg[ArcTan\bigg(\frac{x}{\sqrt{2-x^2}}\bigg)-\frac{\pi}{2}\bigg]\Bigg]
\label{prodbis}
\end{equation}
and
\begin{equation}
F_{P1}=-\frac{\hbar \Omega_{s}}{32 \pi d^3}
\end{equation}
is the force calculated for $\omega_{\tau}=0$ (plasma model) in the low thickness regime. $\Omega_{s}=\Omega_{3}/\sqrt{2}$ is the frequency of the free electron surface plasmon. Here we assumed $\omega_{\tau}<\sqrt{2}\Omega_{3}$, a condition that is commonly verified in real metals. The plot of the 
$g$ function, presented in fig.\ref{fig3} for a large interval of $x$ values, shows that the correction to the force calculated in the ideal plasma model is always negative and gets more significant on increasing 
$\omega_{\tau}$.\\
At high $\Omega_{3}$ values we do not have a simple formula like equation (\ref{prod}) that allows to calculate the force analytically. The plasma wavelength is small and the condition $d\ll\lambda_{p}$ is not satisfied. However we can understand the reason of the increase in the force intensity compared with the plasma model using the following simple argument. The vanishing of the force as $\Omega_{3}$ goes to infinity as due to the exponential factor $e^{-2\xi K_{3}d/c}$ in $Q_{TM}$ and $Q_{TE}$ (see equations (\ref{qtm}) and (\ref{qte})), since the fraction that multiplies the exponential goes to one both in the plasma and Drude model (this corresponds to the condition of reflectivity equal to one). We can write the exponential as $e^{-2 \gamma_{3} d}$, for high $\Omega_{3}$ values we have 
$\gamma_{3}^2\simeq\xi^2\frac{\Omega_{3}}{\xi^2+\xi\omega_{\tau}}$, which shows that for the plasma model the exponential goes to zero faster that for finite
$\omega_{\tau}$ values. This implies that both $Q_{TM}$ and $Q_{TE}$ go to one more slowly as $\omega_{\tau}$ increases, leading to a higher intensity of the force in the Drude model for high $\Omega_{3}$ values.\\
As previously observed \cite{benassi} the occurrence of a maximum is a direct consequence of the retarded nature of the interaction. Indeed formula (\ref{dsmall}), which is obtained by neglecting retardation effects, does not give any maximum as a function of $\Omega_{3}$.\\
\begin{figure}
\centering
\includegraphics[width=8cm,angle=0]{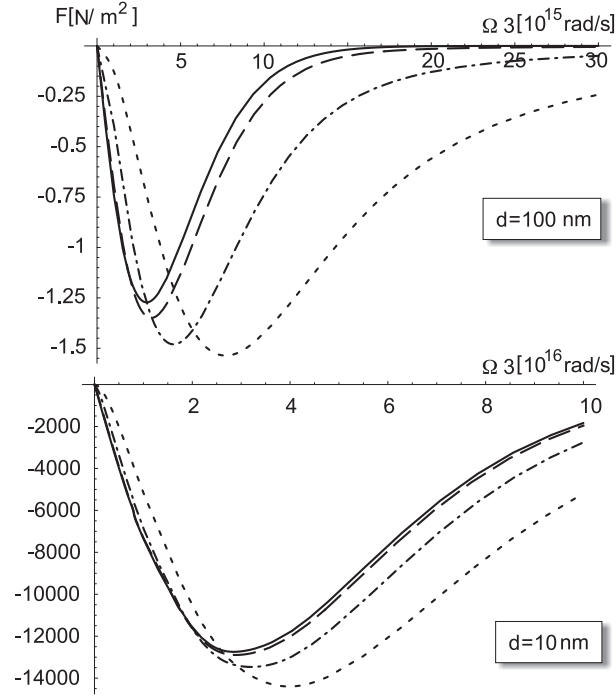}
\caption{\label{fig4}Free standing film. Force on film boundaries as a function of film plasma frequency for different relaxation frequencies. The notation and the $\omega_{\tau}$ values are the same of fig.\ref{fig2}}
\end{figure}
Fig.\ref{fig4} shows the calculated force per unit area in films of different thickness. The shape of the curves and the behaviour as a function of $\omega_{\tau}$ are
basically the same of fig.\ref{fig2} and follow the same trend that has been reported previously for the plasma model \cite{benassi}: the force decreases in intensity and the 
maximum shifts at lower frequencies on increasing $d$.
Again we notice that the inclusion of the relaxation frequency can cause a significant enhancement of the force in the high $\Omega_{3}$ regime.
\begin{figure}
\centering 
\includegraphics[width=8cm,angle=0]{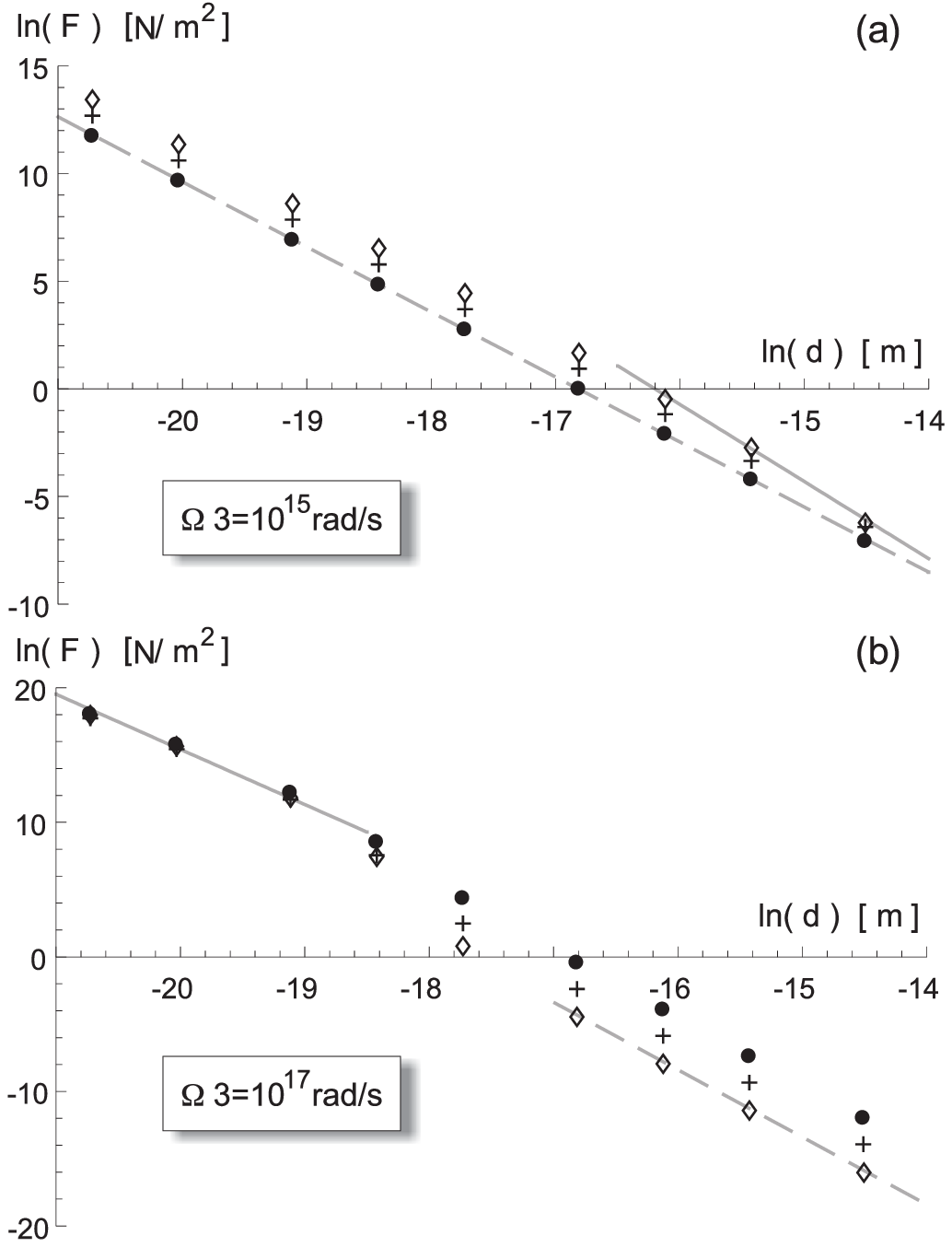}
\caption{\label{fig5}Free standing film. Force as a function of the thickness for different plasma frequency and relaxation frequency values.  $\omega_{\tau}=10^{14}rad/sec$ diamonds, $\omega_{\tau}=10^{15}rad/sec$ crosses, $\omega_{\tau}=5\cdot 10^{15}rad/sec$ dots. Continuous line of figure (a)
is $\sim 1/d^{3.6}$, dashed line of figure (a) is $\sim 1/d^{3}$, continuous line of figure (b) is $\sim 1/d^{4}$ and dashed line of figure (b) is $\sim 1/d^{5}$.}
\end{figure}
Fig.\ref{fig5} displays the plot of the force as a function of $d$ for films of different plasma frequency. It is seen that the force, for relatively low plasma values (fig.\ref{fig5} (a)) behaves as $d^{-3}$ over a large thickness range.
Significant deviations from this behaviour appear at very high relaxation frequencies only. At high plasma frequency 
(fig.\ref{fig5} (b)) different inverse power laws are present in the same interval of thicknesses, in agreement with previous findings \cite{benassi}.\\
The problem of the stability of a free standing film is commonly discussed in terms of competition between surface energy and surface stress and the volume strain energy \cite{streitz,cammarata,ibach}. Surface stresses can arise as a consequence of the different local environment of surface atoms with respect to those in the bulk: if they are tensile stresses they can originate a compressive stress within the film. If the compression is sufficiently large it can cause buckling of the surface with changes the lattice parameters of the outermost of atomic planes.
Treating the film as a continuum one can perform the analysis of the principal strains under the assumption that the film is in a state of biaxial stress \cite{streitz}. For an isotropic system, taking the $z$-axis normal to the surface of the film, one has $\sigma_{xx} =\sigma_{yy} =\sigma$  and $\sigma_{xy} = \sigma_{zx}= \sigma_{yz} = \sigma_{zz} = 0$ and, neglecting quadratic terms, the equilibrium strains are given by:
\begin{equation}
\epsilon_{xx}=\epsilon_{yy}=\epsilon_{\parallel}=\sigma \frac{1-\nu}{Y}\qquad \epsilon_{zz}=\epsilon_{\perp}=-2\sigma\frac{\nu}{Y}
\end{equation}
where $Y$ is the Young modulus and $\nu$ the Poisson's ratio of the metal. In terms of the surface stress $\sigma^{(s)}$
\begin{equation}
\sigma^{(s)}=\gamma+\frac{\partial\gamma}{\partial\epsilon_{\parallel}}
\end{equation}
the equilibrium condition requires
\begin{equation}
\sigma^{(s)}=-\frac{d}{2}\sigma
\end{equation}
so that one can write the strains as
\begin{equation}
\epsilon_{\parallel}=-2\sigma^{(s)}\frac{1-\nu}{Yd}\qquad \epsilon_{\perp}=4\sigma^{(s)}\frac{\nu}{Yd}
\end{equation}
here $d$ is the film thickness and $\gamma$ is the surface energy per unit area. Typical calculated values of the surface stress are of the order of $1\div 5 N/m$ \cite{ibach}. Taking typical values for the elastic parameters one can estimate the strain values for $d=10 nm$ to be of the order of $10^{-3}\div 10^{-4}$.
To include the vacuum fluctuation forces, one has to carry out the same analysis adding to the surface and bulk elastic energy, the vacuum electromagnetic energy of the film. This can be done along the same lines, leading to the following expressions of the strains \cite{landau}
\begin{equation}
\epsilon_{\parallel}=-2\sigma^{(s)}\frac{1-\nu}{Yd}-\nu \frac{F(d)}{Y}\qquad 
\epsilon_{\perp}=4\sigma^{(s)}\frac{\nu}{Yd}+\frac{F(d)}{Y}
\end{equation}
here $F(d)$ is the vacuum electromagnetic force per unit area. As expected by a linear approximation, the strain induced by the surface stress and the vacuum electromagnetic force simply add to determine the strain condition of the system. Even if the contributions have opposite sign, the order of magnitude is different. For $d=10nm$ the maximum value of $F(d)$, as shown in fig.\ref{fig4}, is of the order of $10^{4} N/m^2$, leading to strains of the order of $10^{-5}\div 10^{-6}$, which are too small to compete with the surface stress effects. Therefore, under normal conditions, the electromagnetic force does not contribute significantly to the free standing film morphology. 
For thicknesses of the order of few monolayers, the situation is less clear, since the electromagnetic force is expected to increase at least as $d^{-3}$. However in such cases a dielectric continuum theory is not adequate to represent the film optical properties, since size quantization effects become important. Their inclusion leads to a very different behaviour of the dielectric function and to an expression of the force where the derivative of the film dielectric function with respect to $d$ has to be considered. 
\section{Deposition onto a perfectly reflecting substrate}
\label{dr} 
\begin{figure}
\centering
\includegraphics[width=8cm,angle=0]{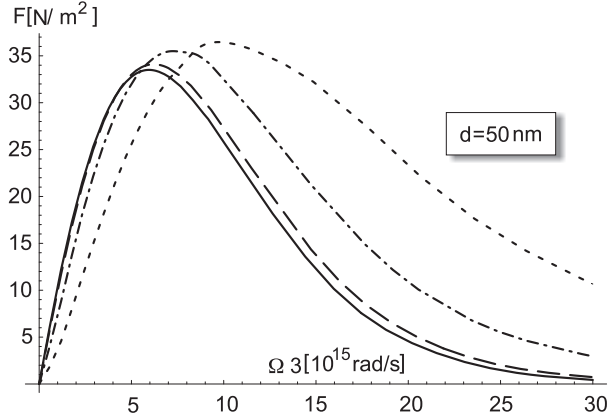}
\caption{\label{fig6}Film deposited on perfectly reflecting substrate. Force on film boundaries as a function of film plasma frequency for different relaxation frequencies.$\omega_{\tau}$ values are the same of fig.\ref{fig2}}
\end{figure}

The situation can be significantly modified if one consider the case of deposited films. Previous work using the plasma model has shown that a metal deposited onto a 
perfectly reflecting substrate is subject to a positive pressure between the boundaries, that tends to increase the film thickness. This force turns out to be approximately double in intensity compared to the force on the free standing film \cite{benassi}.\\
Fig.\ref{fig6} reports the results of the calculation of the force acting on a $50nm$ film deposited on a perfectly reflecting substrate as a function of $\Omega_{3}$
for different $\omega_{\tau}$ values. For comparison, the curve corresponding to the plasma model is also given. One can notice that the force is repulsive and its
intensity is nearly doubled in comparison with the isolated film case. However the shape of the curves is basically the same: the force shows a maximum and a 
long tail at high frequencies. With respect to the plasma model, the inclusion of relaxation leads to an increase of the force in the high $\Omega_{3}$ region and a
decrease at low frequencies.
This behaviour, illustrated in fig.\ref{fig6}, can be understood using the same arguments presented in the previous section. In particular, we notice that in the small $d$ regime the force is simply given by:
\begin{equation}
F=F_{P2}\: f\Bigg(\frac{\omega_{\tau}}{\Omega_{3}}\Bigg)
\label{prod2}
\end{equation}
where
\begin{equation}
f(x)=\frac{1}{\sqrt{1-x^2/2}}\Bigg[1-\frac{2}{\pi}ArcTan\bigg(\frac{x}{\sqrt{2-x^2}}\bigg)\Bigg]
\label{prod2bis}
\end{equation}
and
\begin{equation}
F_{P2}=2\frac{\hbar \Omega_{s}}{32 \pi d^3}=-2F_{P1}
\end{equation}
is the force calculated for the plasma model. The behaviour of the $f(x)$ factor, reported in fig.\ref{fig3}, shows that the force is weakened compared to the plasma case. The strengthening of the force at high $\Omega_{3}$ is basically a consequence of the slower decay of the exponential in equations (\ref{qtm}) and 
(\ref{qte}), as in the case of the free standing film.\\
The $d^{-3}$ dependence is characteristic of the small $d$ regime only and it is appropriate for small plasma frequencies. As $\Omega_{3}$ increase this approximation does not hold any more and, as in the case of the free standing film, different inverse power dependence are needed to represent the force behaviour.\\
The presence of a positive pressure between the boundaries suggests that the force can be useful to stabilize the film. Indeed,
under reasonable assumptions, one can show (see Appendix A) that a metal epitaxial film is stable against any small perturbation when:
\begin{equation}
\frac{\partial^2 E(d)}{\partial d^2}>\frac{\sigma^{4}}{Y^{2}\gamma}
\label{ineq}
\end{equation}
where $\sigma$ is the mismatch lateral stress, $Y$ is the Young modulus of the film, $\gamma$ is the surface energy and $\frac{\partial^2 E(d)}{\partial d^2}$ is
the second derivative of the vacuum fluctuation free energy with respect to the film thickness.\\
\begin{figure}
\centering
\includegraphics[width=8cm,angle=0]{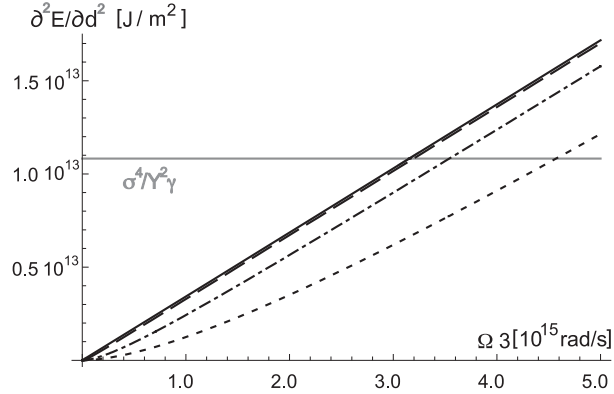}
\caption{\label{fig9}Vacuum energy second derivative plot for the Drude film deposited onto perfectly reflecting substrate. The notation and the $\omega_{\tau}$ values are the same of fig.\ref{fig2}. The film thickness is $d=6nm$.}
\end{figure}
The $d$ value for which the two members of equation (\ref{ineq}) are equal gives the critical thickness $d_{c}$, i.e. the maximum film thickness compatible with a flat surface morphology (see Appendix A).
The critical thicknesses that have been reported in the study of metallic growth are of the order of few nanometers. In this condition and provided $\Omega_{3}$ is not too large, one can use the small $d$ approximation, which corresponds to the van der Waals force description. The energy second derivative turns out to be
\begin{equation}
\frac{\partial^{2}E}{\partial d^{2}}=\frac{3\sqrt{2}\hbar\Omega_{3}}{32\pi d^{4}}f\bigg(\frac{\omega_{\tau}}{\Omega_{3}}\bigg)
\end{equation}
Fig.\ref{fig9} presents curves of $\partial^{2}E/\partial d^{2}$ as a function of  $\Omega_{3}$ for a $6 nm$ film at different relaxation frequencies in the small $d$ limit. The plot allows to determine the range of $\Omega_{3}$ where a flat surface is stable, corresponding to the values of $\partial^{2}E/\partial d^{2}$ above the horizontal line  given by the second member of equation (\ref{ineq}), which has been calculated taking typical values of the parameters representing the elastic and surface properties of the film. 
We used the values $\sigma=500 MPa$, $\gamma=1 J/m^2$, $Y=76GPa$, reported in \cite{zhigang}, changing these values simply shifts the horizontal line in the figure. 
The point where the second derivative curves cross the horizontal line is the minimum plasma frequency that allows for the flat configuration to be stable.
It is seen that, given the film thickness, flat surface stability is possible above a threshold value. On passing from the plasma model, where $\partial^{2}E/\partial d^{2}$ depends linearly on $\Omega_{3}$, to the more realistic Drude description the stability threshold moves to higher $\Omega_{3}$ values.  This indicates that, even adopting a simplified model of the substrate, the conditions of stability depend critically upon the description of the film optical properties.\\
It is common practice to evaluate the vacuum energy contribution using the van der Waals energy. As we noticed before, this is a good approximation only when the film size $d$ is much less that the plasma wavelength. If this condition is not verified, one obtains a rather inaccurate estimate of the critical thickness. To give an example we display in fig.\ref{ctfig1} the $d_c$ values calculated with the full theory, and those obtained with the small $d$ approximation. 
The comparison shows the inadequacy of the van der Waals description in the determination of film stability for a large interval of $\Omega_{3}$ values. Notice that both theories predict the existence of a finite $d_c$ at any $\Omega_{3}$ value. As shown in the next section, this result is peculiar and is a consequence of the perfect reflectivity of the substrate. Relaxation effects play a minor role and lead to differences in the critical values, more significant at large
$\Omega_{3}$.
\begin{figure}
\centering
\includegraphics[width=8cm,angle=0]{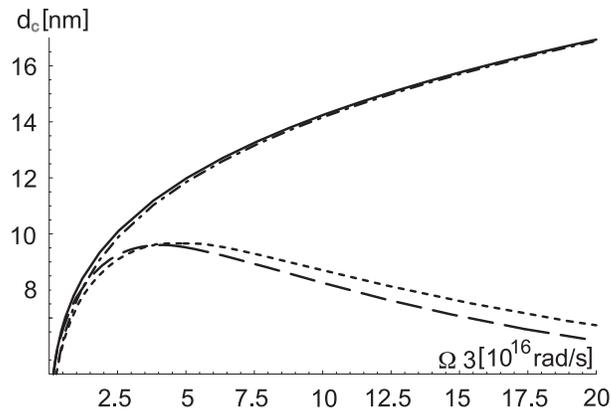}
\caption{\label{ctfig1} Critical thickness for a $6 nm$ film deposited onto a perfectly reflecting substrate. The continuous and dot dashed lines represent respectively the plasma film and Drude films in the small $d$ approximation, $\omega_{\tau}=5\cdot 10^{15}rad/s$. The long dashed and the dashed lines represent again the plasma and Drude films respectively,
but they include also retardation effects.}
\end{figure}
\section{Deposition onto a Drude metal substrate}
\begin{figure}
\centering
\includegraphics[width=8cm,angle=0]{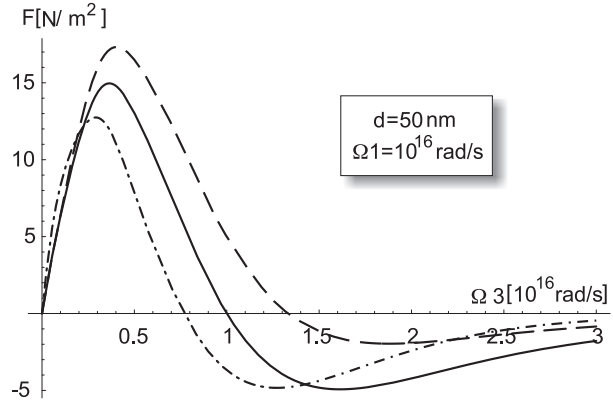}
\caption{\label{fig10}Film deposited onto a Drude substrate. Continuous line, $\omega_{\tau 3}=\omega_{\tau 1}=10^{15}rad/s$; dashed line, $\omega_{\tau 3}=10^{15}rad/s$ and $\omega_{\tau 1}=10^{14}rad/s$; dash dot $\omega_{\tau 3}=10^{14}rad/s$ and $\omega_{\tau 1}=10^{15}rad/s$.}
\end{figure}

\begin{figure}
\centering
\includegraphics[width=8cm,angle=0]{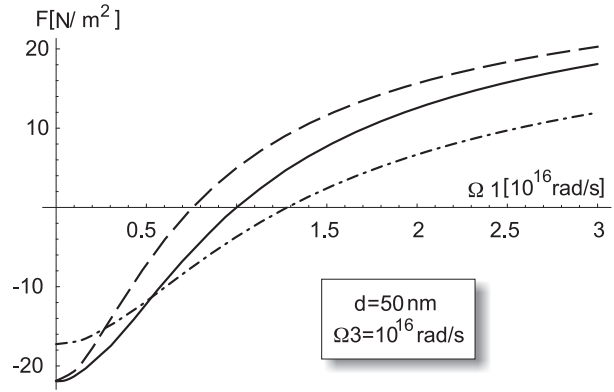}
\caption{\label{fig15}Film deposited onto a Drude substrate. The notation and the $\omega_{\tau}$ values are the same of fig.\ref{fig10}}
\end{figure}

A more realistic description is achieved when the substrate dielectric function is described by a Drude model. 
Fig.\ref{fig10} displays the calculated force per unit area 
for films deposited onto a metallic substrate of fixed plasma frequency $\Omega_{1}$ as a function of $\Omega_{3}$ for different combinations of relaxation frequency
values. The thickness of the film is $50 nm$. The curves are typical of the behaviour that can obtained from the calculations for this kind of systems. The behaviour is
similar to the previously reported results on plasma model \cite{benassi}. The force is repulsive in a limited range of frequencies and becomes negative when $\Omega_{1}$ is
smaller than the film plasma frequencies. This is strictly true when the relaxation frequencies of the substrate and the film are the same. If the substrate relaxation
frequency is lower than the one of the film, the range of frequencies that corresponds to a repulsive force is larger and the force intensity increases, while in the opposite
case the force curve crosses the horizontal axis at lower  $\Omega_{3}$ values.
To understand these findings we notice that the force calculated in the van der Waals, small $d$, regime, when the relaxation frequency is the same for the film and the substrate is given by:
\begin{eqnarray}
\nonumber
F(d)=\frac{\hbar}{16\pi^2d^{3}}\frac{\Omega_{3}^2(\Omega_{1}^2-\Omega_{3}^2)}{\Omega_{s}^2-\bar{\Omega}^2}
\bigg[\frac{1}{\sqrt{4\bar{\Omega}^2-\omega_{\tau}^2}}\bigg(\frac{\pi}{2}-ArcTan\big(\frac{\omega_{\tau}}{\sqrt{4\bar{\Omega}^2-\omega_{\tau}^2}}\big)\bigg)-\\
-\frac{1}{\sqrt{4\Omega_{s}^2-\omega_{\tau}^2}}\bigg(\frac{\pi}{2}-ArcTan\big(\frac{\omega_{\tau}}{\sqrt{4\Omega_{s}^2-\omega_{\tau}^2}}\big)\bigg)\bigg]
\label{lunghissima}
\end{eqnarray}
where $\Omega_{s}=\Omega_{3}/\sqrt{2}$ and $\bar{\Omega}=\sqrt{(\Omega_{1}^2+\Omega_{3}^2)/2}$.
Clearly this approximation predicts a repulsive force if $\Omega_{3}<\Omega_{1}$ and an attractive one if $\Omega_{3}>\Omega_{1}$ in agreement with the result of the full calculation. This conclusion appears to be quite general and independent of the particular model of dielectric function adopted to describe substrate and film dielectric functions. It is consistent with the behaviour of the London dispersion force between dissimilar materials that has been reported in the literature, mainly with reference to ceramic or dielectric materials \cite{israelachvili}.
It has to be noticed that the force may change even by an order of magnitude when the relaxation frequency of the film is modified. This frequency is expected to be different in a film compared to the one measured in the bulk optical properties, due to the contribution of surface scattering and defects. This indicates that the use of the dielectric function of the bulk metal to represent the optical properties of the film may lead to serious errors in the determination of the vacuum force. 
Fig.\ref{fig15} illustrates the behaviour of the force as a function of the substrate plasma frequency for a film with $\Omega_{3}=10^{16} rad/s$, when the relaxation frequency is modified. The curves allow to understand how the force of the film is affected by gradually changing the properties of the substrate. These results like those of fig.\ref{fig10} are quite general since similar curves can be obtained by changing the film thickness and the other parameters. The examples reported in the figure show that the shape and intensity of the force can be significantly modified by varying the optical properties of film and/or substrate. It should be observed that in the real situation of film deposition, the optical properties of the film may be very sensitive to the experimental conditions: both $\Omega_{3}$ and $\omega_{\tau}$ can vary significantly, well above the experimental errors, for films of the same metal prepared by different experimental techniques \cite{pirozhenko}.\\
To discuss the stability problem we notice that, unlike the perfect metal substrate case illustrated in the previous section, deposition onto a real metal does not necessarily give a repulsive force.
\section{Stability in the van der Waals regime}
\begin{figure}
\centering
\includegraphics[width=8cm,angle=0]{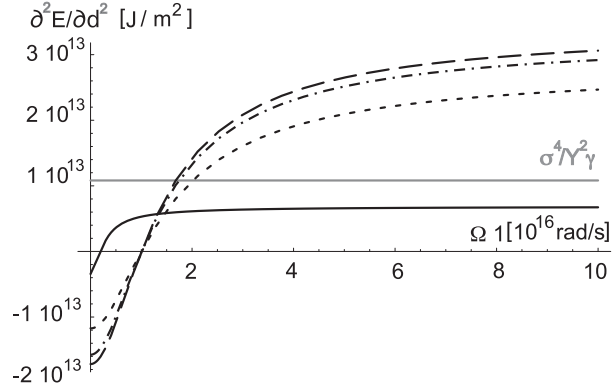}
\caption{\label{fig11}Van der Waals energy second derivative plot. In the case of plasma model we used $\Omega_{3}=2\cdot 10^{15}rad/s$, continuous line and $\Omega_{3}=10^{16}rad/s$, long dashed line. Short dashed line and dot dashed line have been obtained using the Drude model with $\Omega_{3}=10^{16}rad/s$ and  
$\omega_{\tau 3}=10^{15}rad/s$, $\omega_{\tau 1}=10^{14}rad/s$ or $\omega_{\tau 3}=10^{14}rad/s$, $\omega_{\tau 1}=10^{15}rad/s$ respectively. The film thickness is $d=6nm$.}
\end{figure}

\begin{figure}
\centering
\includegraphics[width=8cm,angle=0]{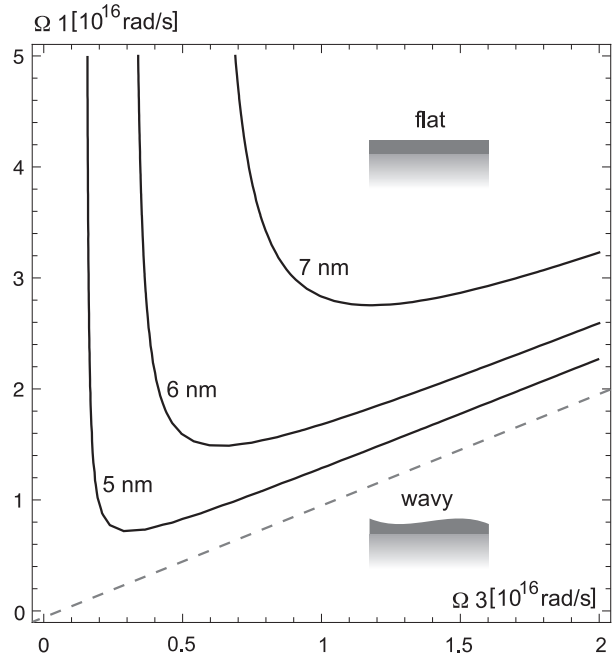}
\caption{\label{fig12}Stability diagram for the plasma film deposited onto a plasma substrate, retardation effects are not included. Different curves for different thicknesses have been represented, above them the film is stable in its wavy morphology, below them the film is stable in its flat morphology.}
\end{figure}

\begin{figure}
\centering
\includegraphics[width=8cm,angle=0]{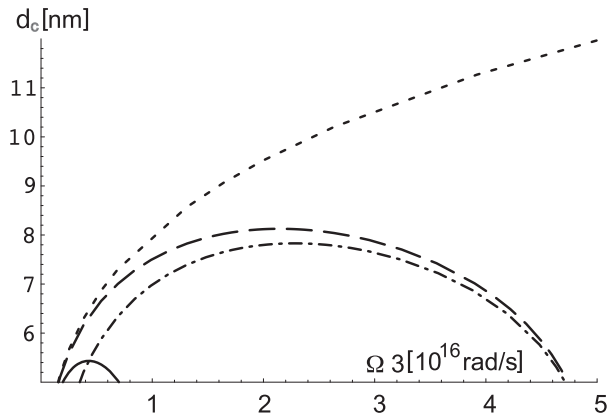}
\caption{\label{ctfig2}Critical thickness. For the plasma model we used $\Omega_{1}=10^{16}rad/s$, continuous line and $\Omega_{1}=5\cdot10^{16}rad/s$, long dashed line. Short dashed line represents the perfectly reflecting substrate $\Omega_{1}\rightarrow\infty$. For the Drude model we used a plasma frequency of $\Omega_{1}=5\cdot10^{16}rad/s$ together with $\omega_{\tau 3}=\omega_{\tau 1}=5\cdot10^{15}rad/s$, dot dashed line.}
\end{figure}

\begin{figure}
\centering
\includegraphics[width=8cm,angle=0]{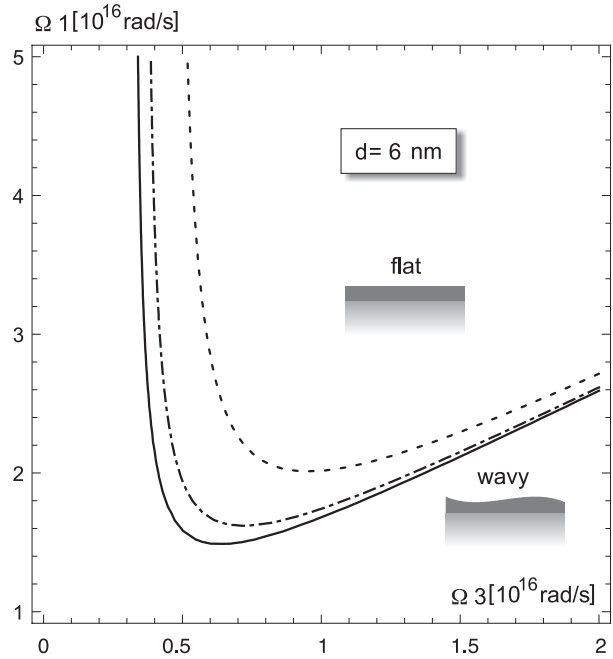}
\caption{\label{fig14}Stability diagram for the Drude film deposited onto a Drude substrate with the same relaxation frequency of the film, retardation effects are not included. Different curves for different relaxation frequencies have been represented, $\omega_{\tau}=5\cdot 10^{15}rad/s$ dashed line, $\omega_{\tau}=10^{15}rad/s$ dot dashed line and $\omega_{\tau}=0$ continuous line.}
\end{figure}

The issue of film stability is currently discussed using empirical forms of the interaction between the film boundaries based on the small $d$ approximation \cite{jiang}. To make reference to the previous work we first present the results obtained under this approximation, to be compared with those obtained including retardation effects that we give in the next section. To illustrate the importance of a realistic description of the optical properties we start discussing the outcomes of calculations using the plasma model i.e. neglecting relaxation effects both in the film and in the substrate. Fig.\ref{fig11} illustrates the behaviour of $\partial^2 E/\partial d^2$ as a function of $\Omega_1$ for films of different plasma frequency and $6nm$ thickness.
For $\Omega_{3}=2\cdot 10^{15}rad/s$ no stability region is achieved since the curve never crosses the horizontal line. This conclusion is obviously dependent on the parameters chosen to represent the elastic and surface properties. 
On the other hand for $\Omega_{3}=10^{16} rad/s$ we find a threshold value above which the flat film is stable.
The difference is caused by the fact that for small $\Omega_{3}$ values even a perfect metal does not provide a force repulsive enough to overcome the instability due to the elastic stress.
Indeed for $\Omega_{1}\rightarrow \infty$ the second derivative of the energy for deposition on a perfect metal in the small $d$ approximation is simply given by:
\begin{equation}
\frac{\partial^2 E(d)}{\partial d^2}=\frac{3 \sqrt{2} \hbar \Omega_{3}}{32 \pi d^4}
\end{equation}
if this quantity is less than $\frac{\sigma^4}{Y^2\gamma}$ one can never achieve a stability region.\\
In the plasma model the only relevant parameters are the plasma frequencies of the film and the substrate. It is useful to represent the results in the form of a stability diagram in the $\Omega_3\Omega_1$ plane, like those given in fig.\ref{fig12}. 
Each curve in the diagram is obtained at fixed film thickness as the locus of the couples $\Omega_{3}\Omega_{1}$ which satisfy the equilibrium condition (\ref{ineq}).
The plane is separated into two areas corresponding to flat surface and rough surface condition respectively.
Notice that upon increasing the film thickness the domain of $\Omega_3\Omega_1$ values that allows for the existence of a flat surface gets narrower. In fig.\ref{ctfig2} we report the calculated critical thickness as a function of $\Omega_3$ for a fixed substrate. Notice that, unlike the perfect metal case, there is a finite range of film plasma values for which the flat surface growth can occur. 
As expected this range reduces as $\Omega_{1}$ decreases. The maximum value of the critical thickness is in correspondence with the maximum intensity of the repulsive force.\\
Turning to the more realistic Drude model we notice in the first place that the values of the second derivative curve are systematically lower that those obtained in the plasma model, as clearly shown in fig.\ref{fig11}. Upon increasing the relaxation frequency the crossing point shifts to higher $\Omega_{3}$ values. Since $\omega_{\tau}$ depends primarily on the quality of the film, this result suggests that the stability condition can be significantly modified by the surface and defect scattering processes that take place in the film. The stability diagrams plotted in fig.\ref{fig14} for a $6 nm$ films illustrate how the inclusion of relaxation processes can vary the range of plasma values allowing for a stable flat surface. We notice from fig.\ref{ctfig2} that the critical thickness values are reduced compared to the plasma model.
\section{Effects of the retardation in the evaluation of deposited film stability}
\begin{figure}
\centering
\includegraphics[width=8cm,angle=0]{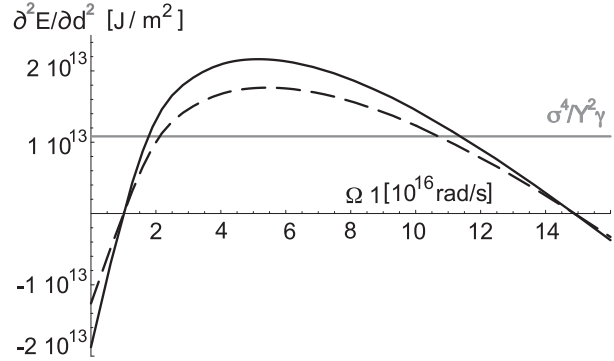}
\caption{\label{fig16}Vacuum energy second derivative plot including retardation effects. Continuous line represents the plasma model with $\Omega_3=10^{16}rad/s$. The Drude model, represented by the dashed line, is obtained with the same plasma frequency and $\omega_{\tau 3}=10^{15}rad/s$ and $\omega_{\tau 1}=10^{14}rad/s$. The critical thickness is $d=6 nm$.}
\end{figure}
The plot of the second derivative curve calculated for the plasma model including retardation as a function of the substrate plasma frequency for a $6 nm$ film, presented in fig.\ref{fig16}, shows that, unlike the van der Waals case, where stability is ensured for all the $\Omega_1$ values above the crossing with the horizontal line, the full theory predicts that the flat surface condition exists for a limited range of $\Omega_1$ values only.
This result puts limitations to stability arguments based on a van der Waals description of the vacuum force, since it shows that the small $d$ approximation is not adequate even at nanometric thicknesses, if the substrate plasma frequency is high enough. The range is smaller when relaxation effects are included into the theory.
\begin{figure}
\centering
\includegraphics[width=8cm,angle=0]{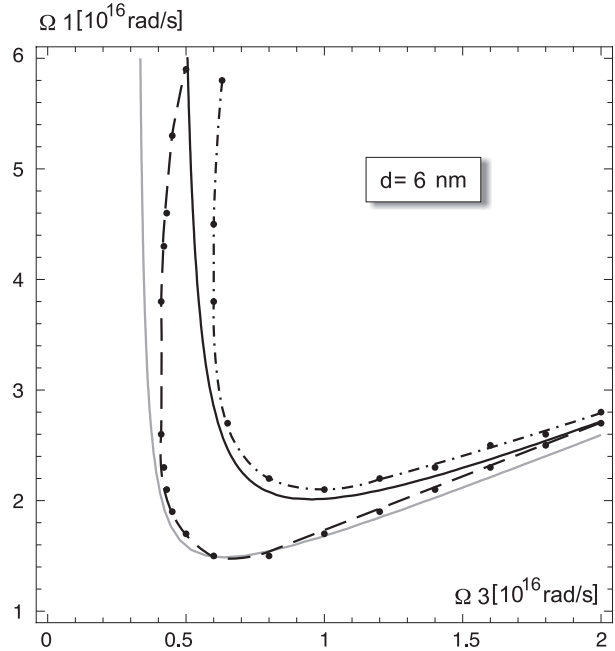}
\caption{\label{fig17}Stability diagram with retardation effects. The dashed and gray continuous lines represent the 
plasma film deposited onto plasma substrate with and without retardation effects respectively.
The dot dashed and black continuous lines represent the Drude film deposited onto Drude substrate with the same relaxation frequency $\omega_{\tau}=5\cdot 10^{15}rad/s$ with and without retardation effects.}
\end{figure}
The stability plots in the $\Omega_3\Omega_1$ plane, given in fig.\ref{fig17}, show the modifications occurring in the range of film-substrate parameters compatible with the flat surface growth, when both retardation and Drude effects are included in the calculations. 
\begin{figure}
\centering
\includegraphics[width=8cm,angle=0]{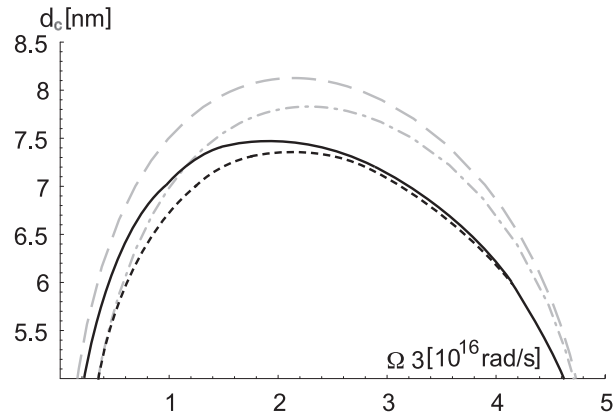}
\caption{\label{ctfig3}Critical thickness including retardation effects. The gray lines represent the plasma and Drude film of fig.\ref{ctfig2} in the small $d$ limit. The continuous and dashed lines represent again the plasma and Drude models respectively, but including retardation effects. The substrate plasma frequency is $\Omega_{1}=5\cdot10^{16}rad/s$.}
\end{figure} 
The reduction in the values of the critical thicknesses for a given $\Omega_{1}$ is illustrated in fig.\ref{ctfig3}.
It is clear from this analysis that the role of the vacuum forces in determining the stability if deposited films depends crucially upon the optical properties of both the film and the substrate. If a reasonably accurate description of the dielectric functions is not available, then it is hard to draw reliable conclusions on  the expected surface morphology. Considerations based on simple empirical expressions of the force based on the van der Waals approximation may lead to overestimate the range of stability, if the conditions that make the small $d$ approximation applicable are not met in the system under consideration.
\section{Conclusions}
The results we have presented illustrate the possibilities that metal films deposited on metal substrates be stabilized by vacuum fluctuation forces. We have shown that there is a variety of situations where these forces may play a role and we have expressed the constraints that have to be satisfied, in order to allow the maintenance of flat surface condition in the film growth, in terms of the parameters entering in the description of the optical properties of both the film and the substrate.\\
The theory is based on a continuum description of the mismatch stress and a dielectric approach to the film and substrate optical properties. The application to a real system requires a detailed knowledge of the film optical properties. The requirement that the force be repulsive as well as its intensity may or may not be satisfied for a film of the same material depending upon the value of the relaxation frequency, the change in electron density that takes place in the pseudomorphic overlayer, the surface scattering, etc. A definite assessment on the role of the vacuum force on a specific system can be made only when these effects are properly accounted in the film dielectric function.\\
The present theory can be improved along two main lines of development. One can improve the continuum approach (i) by adopting a more realistic description of the elastic energy in the study of deposited film stability with the inclusion of strain modifications in the substrate and surface stress effects \cite{mueller,jiang}, which have been neglected in the present paper; (ii) by including interband transition effects into the dielectric function, which are appropriate when transition metal or noble metal are considered \cite{wooten}. One can also extend the theory to deal with the case of metal film onto semiconducting substrate.\\
For very small thickness, i.e. ultra thin films with $d < 1\div 5 nm$, the theory has to be modified to include size quantization effects in the dielectric function \cite{wood}. This is expected to lead to significant modifications in the vacuum force, due to the change in the film dielectric constant caused by the discretization of the electron energy levels. Moreover for ultra thin film the surface energy is expected to depend upon the thickness. Such modifications are consistent with the so called electronic growth model that has been frequently used to describe metal growth onto semiconductor substrate \cite{chiang,czoschke,jia,zhang}. 
\appendix
\section{Macroscopic model for the film stability}
\begin{figure}
\centering
\includegraphics[width=8cm,angle=0]{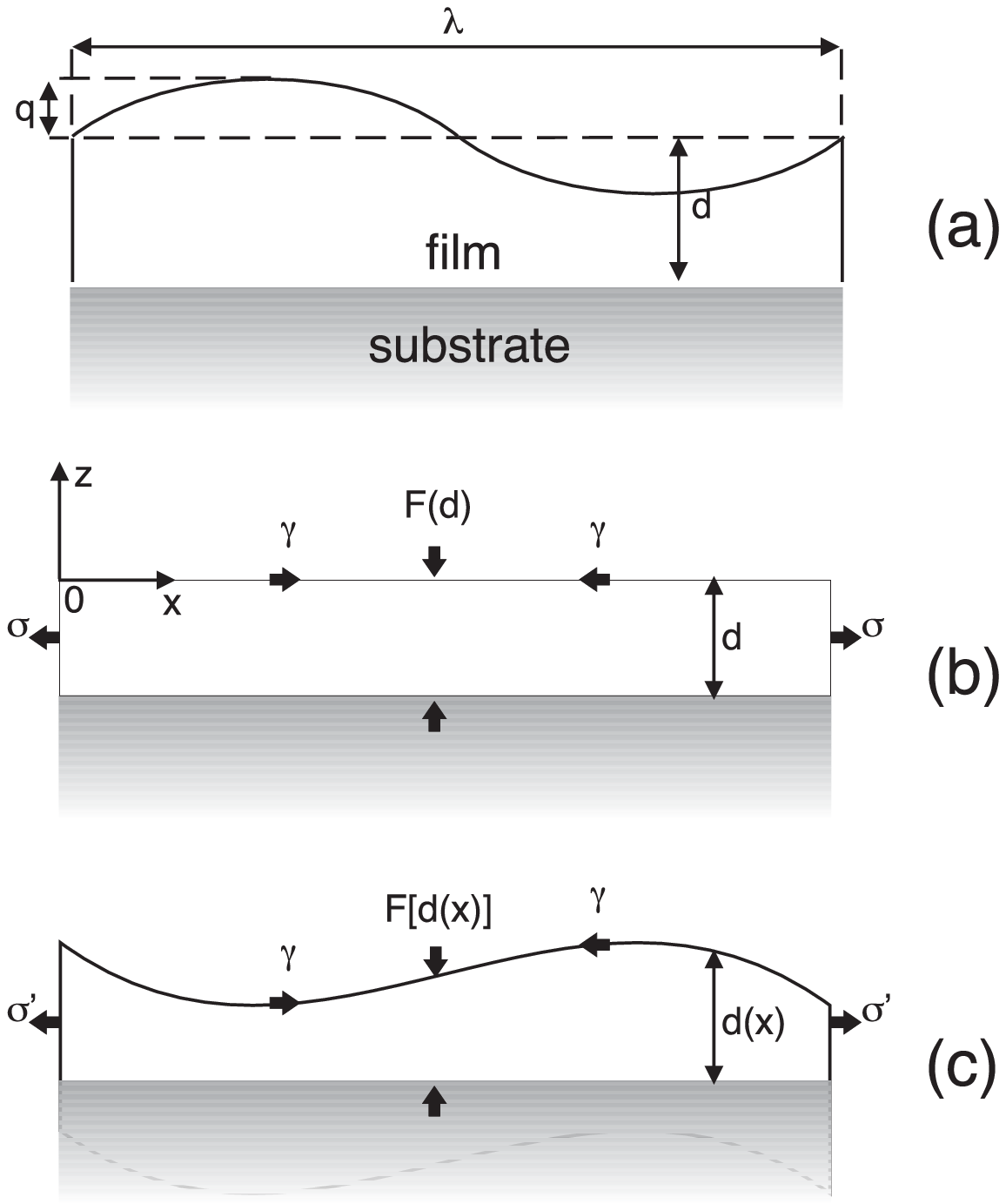}
\caption{\label{appendice}Geometric parameters for the wavy film model and different possible stable configurations.}
\end{figure}

We consider the system sketched in fig. \ref{appendice} where a film of thickness $d$ is deposited on an elastic substrate thick compared to the film. We assume that, under suitable conditions, the film can be grown as a single crystal on the substrate surface. If the film lattice constant is identical to the one of the substrate and the substrate-film interface is perfectly ordered, than the film will grow with zero macroscopic strain. On the other hand, in case of lattice mismatch, and still assuming a perfectly ordered interface, the film will grow continuining the substrate structure and will be strained in the plane of the interface, fig. \ref{appendice} (b). The film will tend to relax this elastic energy through dislocation motion or by changing the shape of the surface through mass transport, fig. \ref{appendice} (c). 
Here we report the analysis where also the second mechanism is present.  
It is generally accepted that at $T=0^{\circ}K$ it is the competition between the elastic energy  and the surface energy that determines the morphological stability of deposited film: indeed when a flat film surface is perturbed into a wave shape, as indicated in the fig. \ref{appendice} (a), elastic energy decreases, but surface energy increases. Surface roughening and island formation are expected to occur as a consequence of this competition, when the flat surface cannot be stable \cite{asaro,grienfield,stolovitz,gao2,gao}.\\ 
However, when the size of the epitaxial film is nanometric, as in micro- nano-scale devices, other forces come into play. Long range dispersion forces, in particular the vacuum fluctuation force between the interface and the film free surface, have been considered as a possible source of stability in the system. 
The analysis is carried out for a sinusoidal surface considered as a perturbation of a reference perfectly planar surface, it is based on perturbation theory and on the assumption that the elastic constant of the film and the substrate are identical and in the absence of a stabilizing influence of the surface stress. For the two dimensional model sketched in fig.\ref{appendice} (a), where $x$ is the surface position and $z=0$ is the reference plane and the sinusoidally wavy shape is described by a cosine function of amplitude q and wavelength $\lambda$:
\begin{equation}
z=d-q\: cos\bigg(\frac{2\pi}{\lambda}x\bigg) 
\end{equation}
one can show \cite{gao} that the stress concentration (stress for unit area and unit length normal to the surface) is related to the stress concentration $\sigma$ for the planar surface configuration by the equations
\begin{eqnarray}
\sigma_{xx}'(x,z)=\sigma\bigg[1 +\frac{4\pi q}{\lambda}\bigg(1-\frac{\pi d}{\lambda}\bigg)e^{-2\pi d/\lambda} cos\bigg(\frac{2 \pi}{\lambda}x\bigg)\bigg]\\
\sigma_{zz}'(x,z)=\sigma\frac{4\pi d q}{\lambda^2}e^{-2\pi d/\lambda}cos\bigg(\frac{2 \pi}{\lambda}x\bigg)\\
\sigma_{xz}'(x,z)=\sigma\frac{2\pi q}{\lambda}\bigg(1-\frac{2 \pi d}{\lambda}\bigg)e^{-2\pi d/\lambda}sin\bigg(\frac{2 \pi}{\lambda}x\bigg)
\end{eqnarray}
which shows that at the surface the stress concentration is minimum in the peaks of the surface wave and maximum in the valleys and that in deeper locations away from the surface the stress is attenuated exponentially with a characteristic length $\lambda/2\pi$. Assuming that $d\ll\lambda$ one has, to the first order, that only the tangential stress contributes to the elastic energy
\begin{equation}
\sigma'(x)=\sigma\bigg[1 +\frac{4\pi q}{\lambda}cos\bigg(\frac{2\pi}{\lambda}x\bigg)\bigg]
\end{equation}
and one can calculate the change in the elastic strain energy density on a wavelength as \cite{zhigang}:
\begin{equation}
\Delta U_{el} = -\sigma^{2}q^{2}\pi/Y
\end{equation}
where $Y$ is the Young modulus. Notice that $\Delta U_{el}$ does not depend upon the wavelength and, being negative, gives rise to film instability for any kind of harmonic perturbation of the surface.\\
The change in the surface energy density can be determined by multiplying the surface tension $\gamma$ by the change in the surface length, which for small amplitude perturbations ($q\ll\lambda)$ leads to the  expression 
\begin{equation}			
\Delta U_{surf}= \gamma q^2 \pi^2/\lambda
\end{equation}
showing that the surface energy increases as the surface is perturbed harmonically. The comparison shows that, independently on the wave amplitude and the film thickness $d$, the flat film surface becomes unstable when:
\begin{equation}
\lambda > \pi Y\gamma/\sigma^2
\end{equation}
the second member of this disequation giving a critical length below which the system is stable.  
Inclusion of the vacuum fluctuation force leads to an additional contribution in the energy density that can be calculated to second order in the surface modifications as:
\begin{equation}
\Delta U_{vac}= \frac{\partial^2 E(d)}{\partial d^2}\Bigg\vert_{d}\frac{q^2 \lambda}{4}
\end{equation}
Adding this term to the previous ones and imposing the equilibrium condition that the total energy change be zero, one arrives at the following conclusions:
\begin{itemize}
\item the second derivative has to be positive, corresponding to a repulsive force on the film;
\item the film is stable provided its thickness $d$ is lower than the critical thickness $d_{c}$ defined through the relation
\begin{equation}
\frac{\partial^2 E(d)}{\partial d^2}\Bigg\vert_{d_{c}}=\sigma^4/\gamma Y^2  
\label{equicond}
\end{equation}
\end{itemize}
For film of very small size, one can assume a van der Waals expression of the vacuum energy 
\begin{equation}
\Delta U_{vac}=-H q^2 \lambda/8\pi d^4
\end{equation}
where $H$ is the Hamaker constant. One obtains the condition
\begin{equation}
d_{c}= \Bigg(\frac{-HY^2\gamma}{2\pi\sigma^4}\Bigg)^{1/4}
\label{ct}
\end{equation}
which shows that the film can be stable only for negative Hamaker constant (repulsive force) and it is inversely proportional to the stress concentration \cite{zhigang,kraft}.\\
Extension of the theory to three dimensions does not modify appreciably the stability condition is a single wave perturbation is considered. In particular equation (\ref{equicond}) is still valid provided that one replaces $Y$ with $Y/(1-\nu^2)$, where $\nu$ is the film Poisson's ratio \cite{mueller}. 
\ack AB thanks \emph{CINECA Consorzio Interuniversitario} ({\tt
www.cineca.it}) for funding his Ph.D. fellowship.
\section*{Reference}
\bibliographystyle{unsrt}

\end{document}